\documentclass[pre,10pt,twocolumn,tightenlines,longbibliography,aps,superscriptaddress,nofootinbib]{revtex4-2}
\usepackage{amsmath}
\usepackage{amssymb}
\usepackage{graphicx}
\usepackage{color}
\usepackage[utf8]{inputenc}
\usepackage[T1]{fontenc}
\usepackage{array}
\usepackage{multirow}
\usepackage[caption=false]{subfig}
\usepackage{hyperref}

\hypersetup{
 colorlinks=true,
 linkcolor=blue,
 filecolor=magenta,
 urlcolor=blue,
 citecolor=blue,
 pdftitle={Finite-time optimal control of forced underdamped harmonic traps and AOUP work extraction},
 pdfpagemode=UseNone,
}

\AtBeginDocument{%
}

\newcommand{\dd}{\mathrm{d}}
\newcommand{\e}{\mathrm{e}}
\newcommand{\Var}{\operatorname{Var}}

\begin{document}

\title{Finite-time minimum-work protocols for underdamped traps driven by active forces}


\author{Kangcan Yao}
\affiliation{Department of Chemical Physics and School of the Gifted Young, University of Science and Technology of China, Hefei 230088, China}

\author{Ruicheng Bao}
\email{ruicheng@g.ecc.u-tokyo.ac.jp}
\affiliation{Department of Physics, Graduate School of Science, The University of Tokyo, Hongo, Bunkyo-ku, Tokyo 113-0033, Japan}

\author{Zhonghuai Hou}
\email{hzhlj@ustc.edu.cn}
\affiliation{Department of Chemical Physics and Hefei National Center for Physical Sciences at the Microscale, University of Science and Technology of China, Hefei 230088, China}

\date{August 16, 2026}

\begin{abstract}
Active systems provide non-equilibrium energy that can, in principle, be harvested using control protocols, making the optimization and reliability of work extraction a fundamental problem in microscopic thermodynamics. Yet how inertia and activity jointly modify the optimal work extraction protocol, the maximum extractable work  its fluctuations, and implementability of protocols remains unclear. Here, we design minimum-work protocols, which maximize work extraction whenever extraction is possible, for an underdamped active particle in a translated harmonic trap. Remarkably, both the minimum mean work and its fluctuations are independent of mass, although inertia modifies the protocol. The results are then applied to an underdamped active Ornstein-Uhlenbeck particle (AOUP), designing an explicit work-extraction criterion  and revealing extracted work's dependence on rotational noise and initial conditions. We further show that inertia strongly affects the difficulty of implementing the optimal protocol, producing a transition-like change between underdamped and overdamped optimal protocols. These findings uncover work extraction mechanisms at the microscale and guide the design of reliable energy-harvesting microscopic devices and robots.
\end{abstract}

\flushbottom
\maketitle

\section{Introduction}\label{sec:intro}

Stochastic thermodynamics extends work, heat, and entropy production to individual fluctuating trajectories, providing a quantitative framework for nonequilibrium processes at the micro- and nanoscale \cite{Bustamante2005,Seifert2012,Ciliberto2017}. Optically trapped colloids have become paradigmatic realizations because their potentials and trajectories can be controlled and measured directly. Active particles add a qualitatively different ingredient: they continuously convert stored or ambient energy into persistent motion \cite{Marchetti2013,Bechinger2016,FodorMarchetti2018,CatesTailleur2015}. Their nonequilibrium force can therefore be both a source of dissipation and a resource for microscopic engines, provided that its temporal structure can be exploited \cite{Speck2016,Mandal2017,FodorJackCates2022}.

Because microscopic operations must be completed in finite time, minimizing their energetic cost is a central problem of stochastic thermodynamics \cite{BlaberSivak2023}. For a passive overdamped particle in a translated harmonic trap, the minimum-work protocol has a linear interior and symmetric jumps of the trap center at the two endpoints \cite{SchmiedlSeifert2007}; numerical variational methods recover the same boundary singularities and extend the calculation to more general potentials \cite{ThenEngel2008}. When inertia is retained, the endpoint jumps persist but ideal delta-peak contributions are additionally required to impose the optimal boundary velocities; remarkably, for trap translation the resulting minimum mean work is unchanged from the overdamped value \cite{GomezMarin2008}, while general underdamped control requires optimization in phase space \cite{SandersBaldovin2024}. Related optimization of Brownian heat-engine cycles has established explicit efficiency-at-maximum-power results \cite{SchmiedlSeifert2008}, while nonequilibrium interactions can generate qualitatively new optimal paths and protocol transitions \cite{Bao2026}.

Several complementary frameworks generalize these exact single-particle results. In the slow-driving regime, thermodynamic length and friction metrics convert dissipation minimization into a geodesic problem \cite{Crooks2007,SivakCrooks2012,Zulkowski2015,Scandi2019}; optimal transport provides finite-time bounds and protocols beyond linear response \cite{Aurell2011,Dechant2022,VanVuSaito2023,ZhongDeWeese2024}. Limited-control formulations extend this connection arbitrarily far from equilibrium \cite{ZhongDeWeese2022}, whereas engineered swift equilibration constructs rapid transformations between prescribed endpoints \cite{Martinez2016,Chupeau2018}. Recent experiments and theory have further identified universal time-reversal structure in optimal protocols and optimal transitions between nonequilibrium steady states \cite{Loos2024,Monter2025}. Finite-time transitions between competing optimal strategies show that the protocol itself can change nonanalytically as duration or constraints are varied \cite{Meibohm2026}.

For active matter, persistence introduces an internal timescale absent from passive control. Response theory predicts how activity changes dissipation and control costs near a reference process \cite{Davis2024}, thermodynamic geometry reveals an activity-dependent optimal transport speed \cite{Wang2025}, and exact control has been obtained for active Brownian particles \cite{Baldovin2023}. For a measured arbitrary time-dependent nonequilibrium force, the optimal moving-trap protocol can use both the initial state and the predictable force history to extract work \cite{Olsen2025}. In moving active traps, open-loop optimization instead gives the passive protocol but larger work fluctuations; conditioning on an initial propulsion measurement changes the protocol and produces a finite-persistence optimum \cite{Schuttler2025}. This closed-loop mechanism supports a minimal active information engine \cite{GarciaMillan2025}, connecting active control to the broader conversion of information into work \cite{SagawaUeda2010,Toyabe2010,Parrondo2015}. Active reservoirs and cyclic protocols likewise demonstrate work extraction at finite power \cite{Krishnamurthy2016,Ekeh2020,Holubec2020,HolubecMarathe2020,BaratoSeifert2017}.

Energetic optimality alone does not guarantee reliable operation. Protocols minimizing mean work and work fluctuations can differ even for a harmonic trap, and their Pareto tradeoff can develop transition-like changes in protocol space \cite{SolonHorowitz2018,Rolandi2023}; analogous distinctions occur among maximum power, efficiency, and stability in microscopic engines \cite{ChenQuan2024,Xu2022}. These results leave open how inertia and a predictable active force jointly determine the finite-time optimal protocol, maximum extractable work, and work fluctuations.

In this paper, we study an underdamped active particle in a translated harmonic trap. In Sec. \ref{sec:model_mean}, we design the finite-time minimum-mean-work protocol by optimizing the mean trajectory and reconstructing the trap center, which shows the optimal mean work is independent of mass. In Sec. \ref{sec:variance}, We compute the thermal work variance of this protocol and show that it is also independent of mass due to the same mode that determines the optimal mean path. In Sec. \ref{sec:aoup}, we illustrate our results through an example of active Ornstein-Uhlenbeck particle (AOUP), and derive the corresponding minimal mean work and variance, and their dependence on rotational diffusion $D_r$ and initial conditions. We futher show that the mass independence of the work statistics does not imply mass-independent implementability: the protocol length and endpoint-impulse strength exhibit a nontrivial tradeoff and a continuous boundary transition when they are optimized jointly.

\begin{figure*}[t]
\centering
\includegraphics[width=0.95\textwidth]{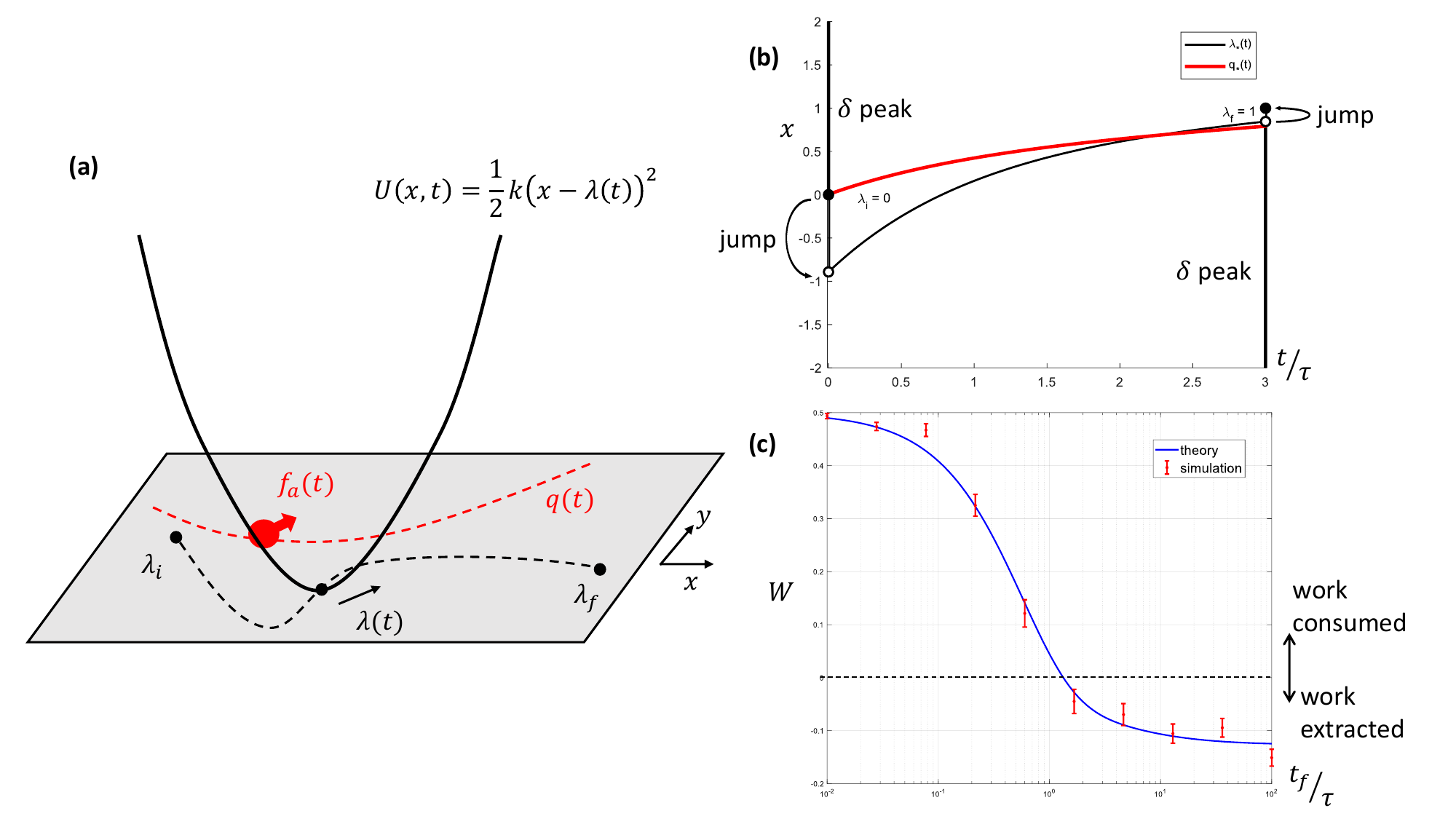}
\caption{Model and one-dimensional benchmark. (a) A Brownian particle (red solid ball) driven by a homogeneous force in two dimensions, $f_a(t)=f_0 e^{-t/\tau}$, is confined by a translated harmonic trap (black solid line) of stiffness $k$, such as an optical tweezer. The mean particle position $q(t)$ is represented by the red dotted line, and the trap center $\lambda(t)$ is represented by the black dotted line. (b) Illustration of the underdamped minimum-work protocol in $x$ direction: the smooth protocol is accompanied by endpoint jumps and ideal $\delta$ peaks, while the mean trajectory remains continuous. The parameters are $m=1$, $\gamma=1$, $k=1$, $\lambda_i=x_i=0$, $\lambda_f=1$, $\tau=1$, $f_0=1$, and $t_f=3$. (c) Comparison of the theoretical and simulated mean work as the protocol duration $t_f$ changes. Positive values correspond to work consumption and negative values to work extraction; the parameters are the same as in (b), except for $t_f$.}
\label{fig:setup}
\end{figure*}

\section{Model and minimum-mean-work protocol}\label{sec:model_mean}
We consider a particle in a translated harmonic trap,
$V(x,\lambda)=k[x-\lambda(t)]^2/2$, whose center is moved from
$\lambda_i$ to $\lambda_f$ in a fixed time $t_f$. The active force available
to the controller is denoted by $f_a(t)$. It may be an imposed reproducible
drive or, in the active-particle application below, a conditional mean
self-propulsion inferred from measurement information. This setting is
sketched in Fig.~\ref{fig:setup}(a). The particle is initially located at $x(0)=q_i$.
The underdamped Langevin equation is
\begin{equation}
 m\ddot x+\gamma\dot x+k\bigl[x-\lambda(t)\bigr]
 =f_a(t)+\sqrt{2\gamma k_B T}\,\xi(t),
 \label{eq:langevin}
\end{equation}
where the thermal fluctuations are modeled by Gaussian white noise with $\langle\xi(t)\rangle=0$ and
$\langle\xi(t)\xi(t')\rangle=\delta(t-t')$. The controller work is
\begin{equation}
 \langle W\rangle=
 k\int_0^{t_f}\dot\lambda(t)\langle\lambda(t)-x(t)\rangle\dd t ,
 \label{eq:work}
\end{equation}
where positive work means energy consumption and negative work means extraction.

Let $q(t)=\langle x(t)\rangle$ be the mean trajectory of the particle. Its dynamics obey
\begin{equation}
 m\ddot q+\gamma\dot q+k\bigl[q-\lambda(t)\bigr]=f_a(t).
 \label{eq:mean_dyn}
\end{equation}
The optimization is simplest if the mean trajectory, rather than the trap
center, is used as the variational variable. Eliminating $\lambda(t)$ gives
the local work functional
\begin{equation}
 \langle W\rangle=
 \left[\frac{k}{2}(\lambda-q)^2+\frac{m}{2}\dot q^{\,2}\right]_0^{t_f}
 +\int_0^{t_f}\left[\gamma\dot q^{\,2}-f_a(t)\dot q\right]\dd t .
 \label{eq:mean_functional}
\end{equation}
The integral contains the dissipative cost $\gamma\dot q^{\,2}$ and the work
harvested from the active drive, $-f_a\dot q$. The mass appears only through
boundary terms, which is why it affects the protocol but not the final
minimum-work value.
The Euler--Lagrange equation, $2\gamma\ddot q-\dot f_a=0$, gives
\begin{equation}
 \dot q_*(t)=g(t),\qquad
 g(t)=\phi+\frac{f_a(t)}{2\gamma},
 \label{eq:gdef}
\end{equation}
where
\begin{equation}
 \phi=\frac{k}{2\gamma+k t_f}
 \left[\lambda_f-q_i-\frac{1}{2\gamma}\int_0^{t_f} f_a(s)\dd s\right],
 \label{eq:phi}
\end{equation}
and $q_i=q(0)$. Hence
\begin{equation}
 q_*(t)=q_i+\phi t+\frac{1}{2\gamma}\int_0^t f_a(s)\dd s .
 \label{eq:qstar}
\end{equation}

Solving Eq.~\eqref{eq:mean_dyn} for the trap center gives the smooth part of the optimal protocol,
\begin{equation}
 \lambda_*^{\rm s}(t)=q_*(t)+\frac{\gamma g(t)}{k}
 -\frac{f_a(t)}{k}+\frac{m\dot g(t)}{k},
 \qquad 0<t<t_f .
 \label{eq:lambdastar_explicit}
\end{equation}
The force-dependent terms reproduce the overdamped active-force correction of
Ref.~\cite{Olsen2025}, while $m\dot g/k$ reflects how the inertia affects the optimal protocol.

The full underdamped protocol can be viewed as the
open-interval protocol together with the boundary operations of
underdamped finite-time control and can be written as \cite{GomezMarin2008}:
\begin{align}
 \lambda_*(t)={}&\lambda_i+
 \bigl[\lambda_*^{\rm s}(t)-\lambda_i\bigr]\Theta(t)
 +\bigl[\lambda_f-\lambda_*^{\rm s}(t)\bigr]\Theta(t-t_f)
 \notag\\
 &+\frac{m g(0)}{k}\delta(t)
 -\frac{m g(t_f)}{k}\delta(t-t_f).
 \label{eq:full_protocol}
\end{align}
The first line represents the endpoint jumps of the trap center, and the
second line represents ideal endpoint impulses that impose the optimal
one-sided mean velocities. In what follows, only the resulting formulas are
used; Appendix~\ref{app:mean} records the detailed derivation. The
structure of these jumps and impulses is shown schematically in
Fig.~\ref{fig:setup}(b).

The corresponding minimum mean work is
\begin{align}
 \langle W\rangle_*={}&\frac{\gamma k}{2\gamma+k t_f}
 \left[\lambda_f-q_i
 -\frac{1}{2\gamma}\int_0^{t_f} f_a(s)\dd s\right]^2 \notag\\
 &-\frac{k}{2}(\lambda_i-q_i)^2
 -\frac{1}{4\gamma}\int_0^{t_f}f_a(t)^2\dd t .
 \label{eq:Wstar_compact}
\end{align}
The first term penalizes the mismatch between the desired transport and the
active displacement accumulated during the protocol, while the last term is
the active-force resource available for extraction. The value of the optimum
is independent of $m$ as expected, whereas the protocol realizing it is not.
Figure~\ref{fig:setup}(c) illustrates this criterion in a finite-time
benchmark: increasing the protocol duration reduces dissipative cost until
the active-force contribution can make the mean work negative, thereby enabling work extraction.
A longer duration reduces the mismatch and allows more work induced by the active force to be harvested.

\section{Work fluctuations of the minimum-mean-work protocol}\label{sec:variance}

Directly computing the work variance from the full underdamped protocol $\lambda_*(t)$ is cumbersome.
We first separate the deterministic and fluctuating parts of the particle trajectory by writing
\begin{equation}
 x(t)=q_*(t)+\eta(t),
\end{equation}
with $x(0)=q_i$ and $\dot x(0)=0$.
The fluctuation coordinate obeys the homogeneous underdamped Langevin equation
\begin{equation}
 m\ddot\eta+\gamma\dot\eta+k\eta
 =\sqrt{2\gamma k_B T}\,\xi(t),
 \qquad \eta(0)=\dot\eta(0)=0.
 \label{eq:eta}
\end{equation}
Let $h(t)$ be the causal Green function of $m\partial_t^2+\gamma\partial_t+k$. The fluctuating work is
\begin{equation}
 \delta W=-k\int_0^{t_f}\dot\lambda_*(t)\eta(t)\dd t
 =-\sqrt{2\gamma k_B T}\int_0^{t_f}\Psi(t)\xi(t)\dd t,
 \label{eq:deltaW}
\end{equation}
where the backward kernel is
\begin{equation}
 \Psi(t)=\int_t^{t_f}k\dot\lambda_*(s)h(s-t)\dd s .
 \label{eq:Psi_def}
\end{equation}
Using the distributional protocol in Eq.~\eqref{eq:full_protocol}, the
adjoint calculation in Appendix~\ref{app:kernel} gives
\begin{equation}
 \Psi(t)=g(t)=\phi+\frac{f_a(t)}{2\gamma}.
 \label{eq:Psi_g}
\end{equation}
Therefore the conditional work distribution is Gaussian and
\begin{equation}
 \Var_{\rm cond}(W_*)=
 2\gamma k_B T\int_0^{t_f}
 \left[\phi+\frac{f_a(t)}{2\gamma}\right]^2\dd t .
 \label{eq:var_star}
\end{equation}
The mass does not appear in Eq.~\eqref{eq:var_star}. It is surprising that optimal mean work and its variance are both independent of mass. This does not mean
that inertia is dynamically or experimentally irrelevant. The inertial Green function
and the boundary operations combine so that the stochastic work is projected
onto the same mode $g(t)$ that determines the optimal mean path, but the trap
trajectory and its singular endpoint operations remain strongly mass dependent.
Thus identical work statistics can conceal substantially different control
requirements, as quantified below for the AOUP protocol.

\section{Active-force application, work extraction, and implementability}\label{sec:aoup}
We then apply the preceding results to an active particle. Above, the external
force is deterministic. In the active application, the active force is stochastic,
so we instead use its conditional mean given the initially available information.

\begin{equation}
 \boldsymbol F_a(t)=\langle\boldsymbol f_a(t)\mid \mathcal I\rangle,
\end{equation}
where $\mathcal I$ denotes the information available to the controller, such
as an initial active-orientation measurement.
For a one-dimensional AOUP force,
\begin{equation}
 \tau_a\dot f_a=-f_a+\sqrt{2D_a}\,\xi_a(t),
 \label{eq:aoup_force}
\end{equation}
where $\langle \xi_a(t) \xi_a(t') \rangle=\delta(t-t')$.
An exact measurement $f_a(0)=f_i$ gives
\begin{equation}
 F_a(t)=f_i\e^{-t/\tau_a}.
 \label{eq:aoup_mean}
\end{equation}
The needed integrals are
\begin{align}
 \int_0^{t_f}F_a(t)\dd t&=f_i\tau_a\bigl(1-\e^{-t_f/\tau_a}\bigr),
 \notag\\
 \int_0^{t_f}F_a(t)^2\dd t&=
 \frac{f_i^2\tau_a}{2}\bigl(1-\e^{-2t_f/\tau_a}\bigr).
 \label{eq:aoup_integrals}
\end{align}
These formulas recover the finite-persistence work-extraction mechanism
discussed for active particles in moving traps \cite{Schuttler2025}, while
retaining the underdamped protocol and the thermal variance structure derived
above.

The preceding formulas can also be applied componentwise to vector traps.
For a two-dimensional particle,
\begin{equation}
 m\ddot{\boldsymbol x}+\gamma\dot{\boldsymbol x}
 +k(\boldsymbol x-\boldsymbol\lambda)
 =\boldsymbol F_a(t)+\sqrt{2\gamma k_B T}\,\boldsymbol\xi(t),
 \label{eq:vector}
\end{equation}
the optimal velocity and integration constant are
\begin{align}
 \boldsymbol g(t)&=\boldsymbol\phi+\frac{\boldsymbol F_a(t)}{2\gamma},
 \label{eq:vector_g}\\
 \boldsymbol\phi&=\frac{k}{2\gamma+k t_f}
 \left[\boldsymbol\lambda_f-\boldsymbol q_i
 -\frac{1}{2\gamma}\int_0^{t_f}\boldsymbol F_a(s)\dd s\right],
 \label{eq:vector_phi}
\end{align}
and the minimum mean work becomes
\begin{align}
 \langle W\rangle_*={}&
 \frac{\gamma k}{2\gamma+k t_f}
 \left|\boldsymbol\lambda_f-\boldsymbol q_i
 -\frac{1}{2\gamma}\int_0^{t_f}\boldsymbol F_a(s)\dd s\right|^2
 \notag\\
 &-\frac{k}{2}\left|\boldsymbol\lambda_i-\boldsymbol q_i\right|^2
 -\frac{1}{4\gamma}\int_0^{t_f}|\boldsymbol F_a(t)|^2\dd t .
 \label{eq:vector_W}
\end{align}
Equation~\eqref{eq:vector_W} is the finite-time extraction criterion. The
first term penalizes mismatch between the target displacement and the
time-integrated conditional force. The last term is nonpositive and measures
the predictable active-force contribution that can be harvested.

\begin{figure*}[t]
\centering
\includegraphics[width=0.98\textwidth]{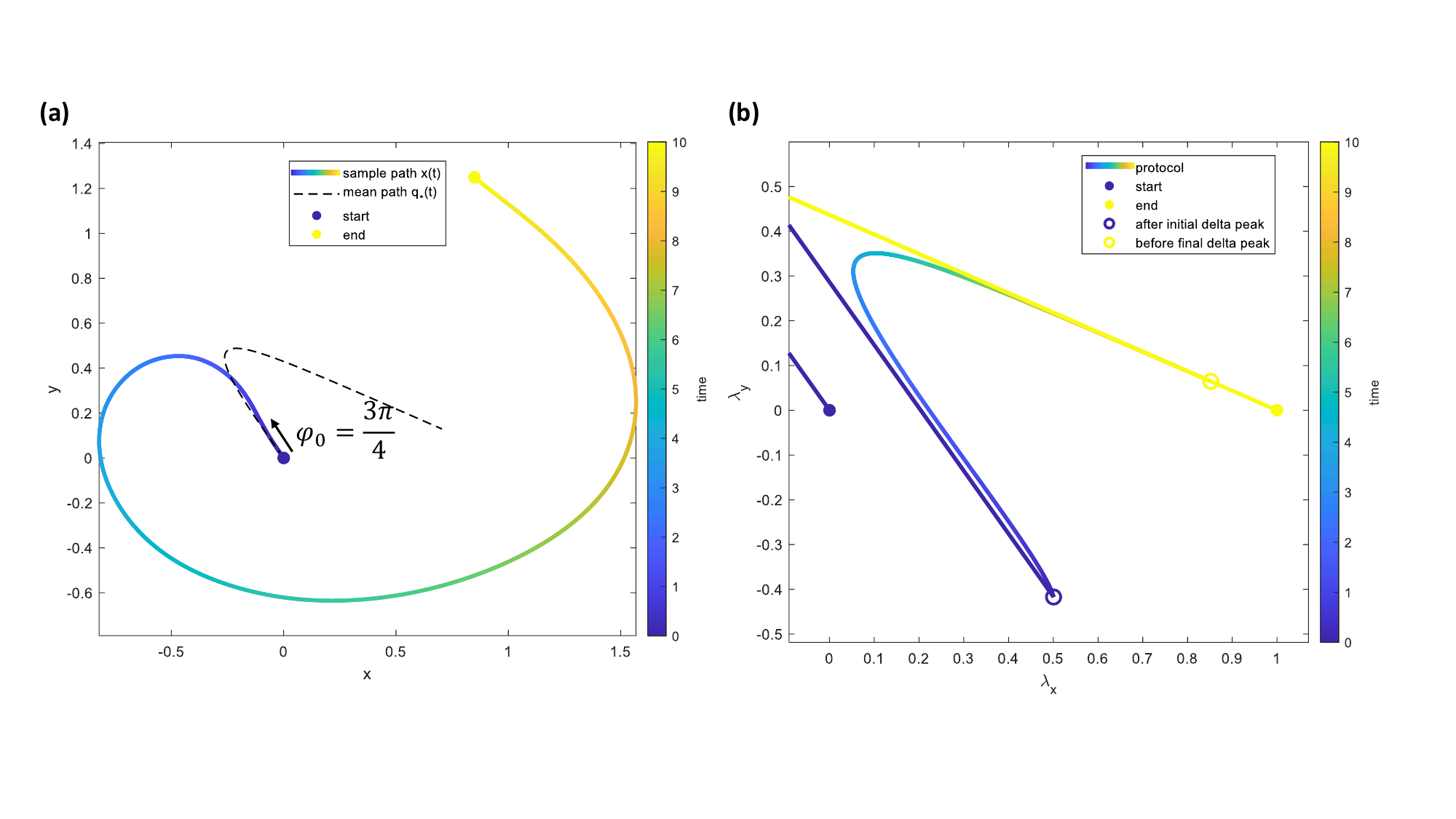}
\caption{Two-dimensional AOUP optimal protocol and mean trajectory. (a) A sample trajectory colored by time, together with the optimal mean path for the initial orientation $\varphi_0=3\pi/4$. The active force is defined in Eq.~\eqref{eq:orientation}. (b) The corresponding trap protocol in the $(\lambda_x,\lambda_y)$ plane, including the start and end points and the one-sided points associated with the endpoint impulses. The parameters are $m=1$, $\gamma=1$, $k=1$, $\lambda_i=x_i=(0,0)$, $\lambda_f=(1,0)$, $\tau_R=2$, $D_r=1$, $f_0=1$, and $t_f=10$.}
\label{fig:trajectory_protocol}
\end{figure*}

Figure~\ref{fig:trajectory_protocol} gives a representative two-dimensional
realization of this vector construction. The sample path bends strongly under
the active orientation, while the trap protocol compensates for the
conditioned active force and includes the endpoint operations inherited from
the underdamped solution.

\begin{figure*}[t]
\centering
\includegraphics[width=0.98\textwidth]{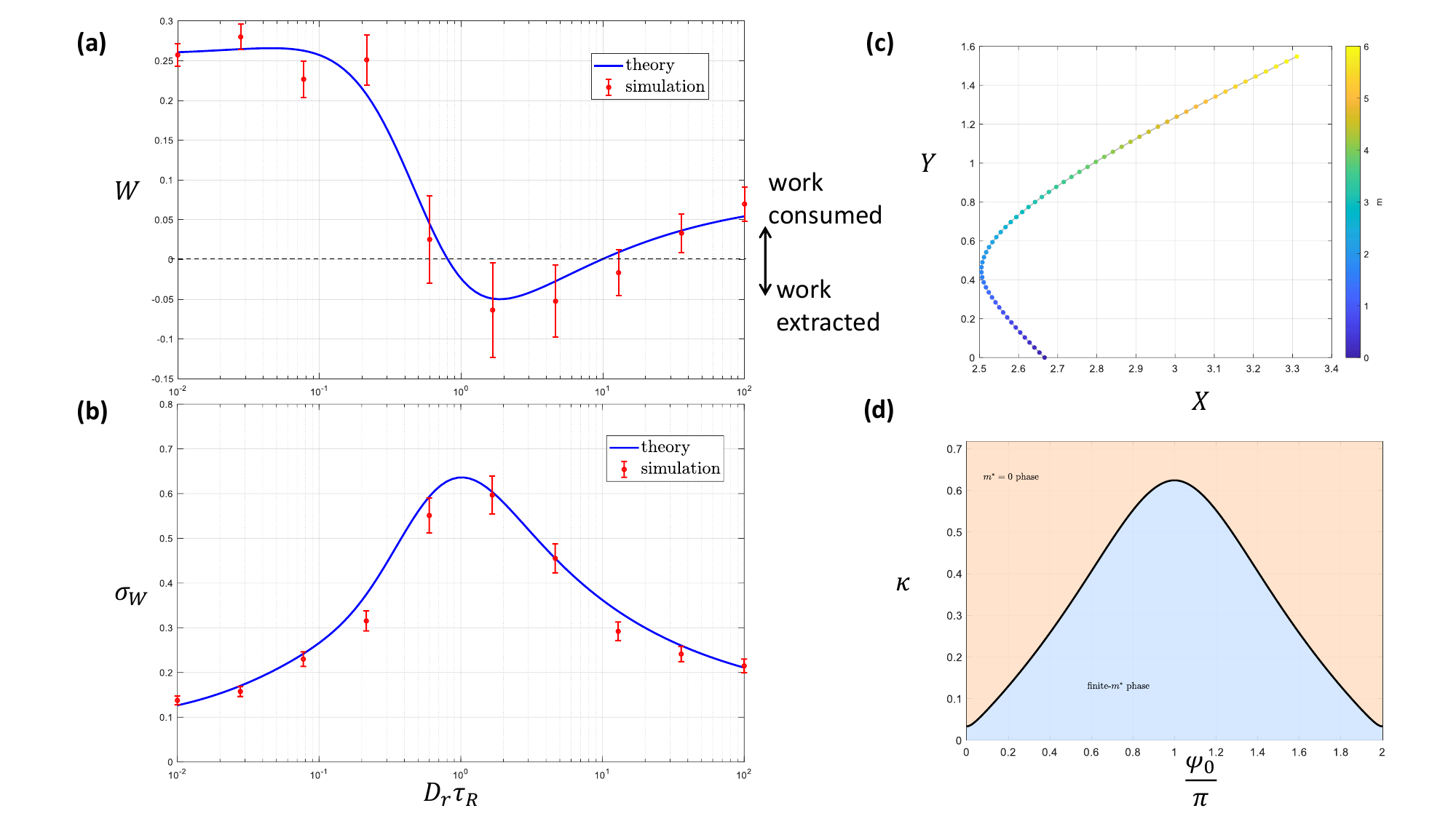}
\caption{AOUP work statistics and mass-dependent implementability. (a) The theoretical work $W_{\rm th}$ from Eq.~\eqref{eq:W_orientation} and the simulated work $W_{\rm ph}$ as functions of $D_r\tau_R$. Positive work corresponds to work consumption, while negative work corresponds to work extraction. (b) The corresponding work standard deviation. The theoretical result from Eq.~\eqref{eq:active_var} and the simulations both show an enhancement at intermediate $D_r\tau_R$. The parameters in (a) and (b) are $m=1$, $\gamma=1$, $k=1$, $\boldsymbol\lambda_i=\boldsymbol q_i=(0,0)$, $\boldsymbol\lambda_f=(1,0)$, $\tau_R=2$, $f_0=1$, $\varphi_0=3\pi/4$, $t_f=10$, $k_B=1$, and $T=0.1$. (c) Mass-parametrized tradeoff between the regular-protocol length $X$ and the normalized total endpoint-impulse strength $Y$ defined in Eq.~\eqref{eq:implementation_costs}; color denotes $m$. The remaining parameters are the same as in (a) and (b), with $m$ scanned from $0$ to $6$. (d) Control-implementation diagram obtained by minimizing $J_\kappa=X+\kappa Y$. The black curve is the critical weight $\kappa_c(\varphi_0)$ separating an optimum at finite $m^*$ (blue) from the boundary optimum $m^*=0$ (orange).}
\label{fig:aoup}
\end{figure*}

For a two-dimensional AOUP model,
\begin{equation}
 \boldsymbol f_a(t)=f_0[\cos\varphi(t),\sin\varphi(t)],\qquad
 \dot\varphi=\Omega,
 \label{eq:orientation}
\end{equation}
with colored angular velocity
\begin{equation}
 \dot\Omega=-\frac{\Omega}{\tau_R}
 +\frac{\sqrt{2D_r}}{\tau_R}\xi(t).
 \label{eq:omega_ou}
\end{equation}
Here the angular fluctuations are modeled by Gaussian white noise with $\langle \xi(t) \xi(t') \rangle=\delta(t-t')$,
$\tau_R$ is the rotational relaxation time, and $D_r$ is the rotational diffusion coefficient.
Conditioned on the initial direction $\varphi_0$, and initial angular velocity $\Omega_0$ 
obeying a distribution $\mathcal P(\Omega_0)=\sqrt{\tau_R/(2\pi D_r)}\exp(-\tau_R\Omega_0^2/(2D_r))$,
the mean active force is
\begin{equation}
 \boldsymbol F_a(t|\varphi_0)=
 f_0\e^{-D_r[t-\tau_R(1-\e^{-t/\tau_R})]}
 [\cos\varphi_0,\sin\varphi_0].
 \label{eq:mean_orientation}
\end{equation}
Define
\begin{equation}
 B(t)=t-\tau_R(1-\e^{-t/\tau_R}),\qquad
 R_n(D_r)=\int_0^{t_f}\e^{-nD_r B(t)}\dd t ,
 \label{eq:Rn}
\end{equation}
and $\boldsymbol u_0=(\cos\varphi_0,\sin\varphi_0)$. Then
$\int_0^{t_f}\boldsymbol F_a(t)\dd t=f_0R_1(D_r)\boldsymbol u_0$ and
$\int_0^{t_f}|\boldsymbol F_a(t)|^2\dd t=f_0^2R_2(D_r)$. Substituting
Eq.~\eqref{eq:mean_orientation} into Eq.~\eqref{eq:vector_W} gives
\begin{align}
 W_{\rm th}={}&
 \frac{\gamma k}{2\gamma+k t_f}
 \left|\boldsymbol\lambda_f-\boldsymbol q_i
 -\frac{f_0R_1(D_r)}{2\gamma}\boldsymbol u_0\right|^2
 \notag\\
 &-\frac{k}{2}\left|\boldsymbol\lambda_i-\boldsymbol q_i\right|^2
 -\frac{f_0^2R_2(D_r)}{4\gamma}.
 \label{eq:W_orientation}
\end{align}
This expression gives the theoretical curve in Fig.~\ref{fig:aoup}(a), showing
a nonmonotonic dependence on $D_r\tau_R$ for the misaligned initial orientation.

The fluctuating active force around the conditional mean gives an additional
contribution to the work variance,
\begin{align}
 \Var(W_*)={}&2\gamma k_B T\int_0^{t_f}|\boldsymbol g(t)|^2\dd t
 \notag\\
 &+\int_0^{t_f}\!\int_0^{t_f}
 \boldsymbol g(t)\cdot C_a(t,t')\boldsymbol g(t')
 \dd t\dd t',
 \label{eq:active_var}
\end{align}
where
$C_a(t,t')=\langle\delta\boldsymbol f_a(t)\delta\boldsymbol f_a(t')\rangle$.
The first term is the thermal contribution derived in
Sec.~\ref{sec:variance}, while the second term is the contribution from residual
active-force correlations. Equation~\eqref{eq:active_var}
is used for Fig.~\ref{fig:aoup}(b).

Although Eqs.~\eqref{eq:W_orientation} and \eqref{eq:active_var} contain no
mass, implementing their common optimum does. We characterize this distinction
using two geometric control costs. Let $X$ be the total path length of the
regular part of the trap protocol, including its finite endpoint jumps but
excluding the ideal Dirac peaks, and let
\begin{equation}
 Y=\frac{|\boldsymbol A_0|+|\boldsymbol A_f|}{\tau_R},
 \qquad
 \boldsymbol A_0=\frac{m}{k}\boldsymbol g(0),\quad
 \boldsymbol A_f=-\frac{m}{k}\boldsymbol g(t_f),
 \label{eq:implementation_costs}
\end{equation}
measure the normalized total strength of the two endpoint impulses. The
normalization by $\tau_R$ gives $X$ and $Y$ the same dimensions. These costs
probe complementary experimental limitations: $X$ measures how far the trap
must travel along the regular protocol, whereas $Y$ measures the impulsive
control authority required at the boundaries.

Figure~\ref{fig:aoup}(c) makes the separation between energetic performance
and implementability explicit. Increasing $m$ initially shortens the regular
trap path because the inertial correction $m\dot{\boldsymbol g}/k$ partly
reshapes the active-force contribution. At the same time, however, the total
impulse cost $Y$ grows linearly with $m$. Beyond the minimum of $X$, both costs
increase, so that part of the mass-parametrized curve is dominated. Hence a
particle can have exactly the same minimum mean work and work variance as a
lighter particle while requiring a substantially longer trap trajectory and
larger endpoint impulses to attain them.

To express the competition between the two implementation costs, we minimize
$J_\kappa=X+\kappa Y$, where $\kappa\geq0$ specifies the relative penalty on
impulsive control. For small $\kappa$, the path-shortening benefit selects a
finite optimal mass $m^*>0$. Above a critical $\kappa_c$, impulse suppression
dominates and the optimum reaches the overdamped boundary $m^*=0$.
Figure~\ref{fig:aoup}(d) shows that this boundary depends strongly on the
initial active orientation and is largest near $\varphi_0=\pi$, where the
active force initially opposes the target displacement most strongly. For the
parameters of Fig.~\ref{fig:aoup}(c), $\kappa_c=0.519$. The optimum approaches
zero continuously, so this is a transition-like change of a constrained
control optimum rather than a thermodynamic phase transition. Appendix~\ref{app:implementation}
derives the convexity of $X(m)$, the critical-weight condition, and the
near-critical scaling.

\begin{figure*}[t]
\centering
\includegraphics[width=0.98\textwidth]{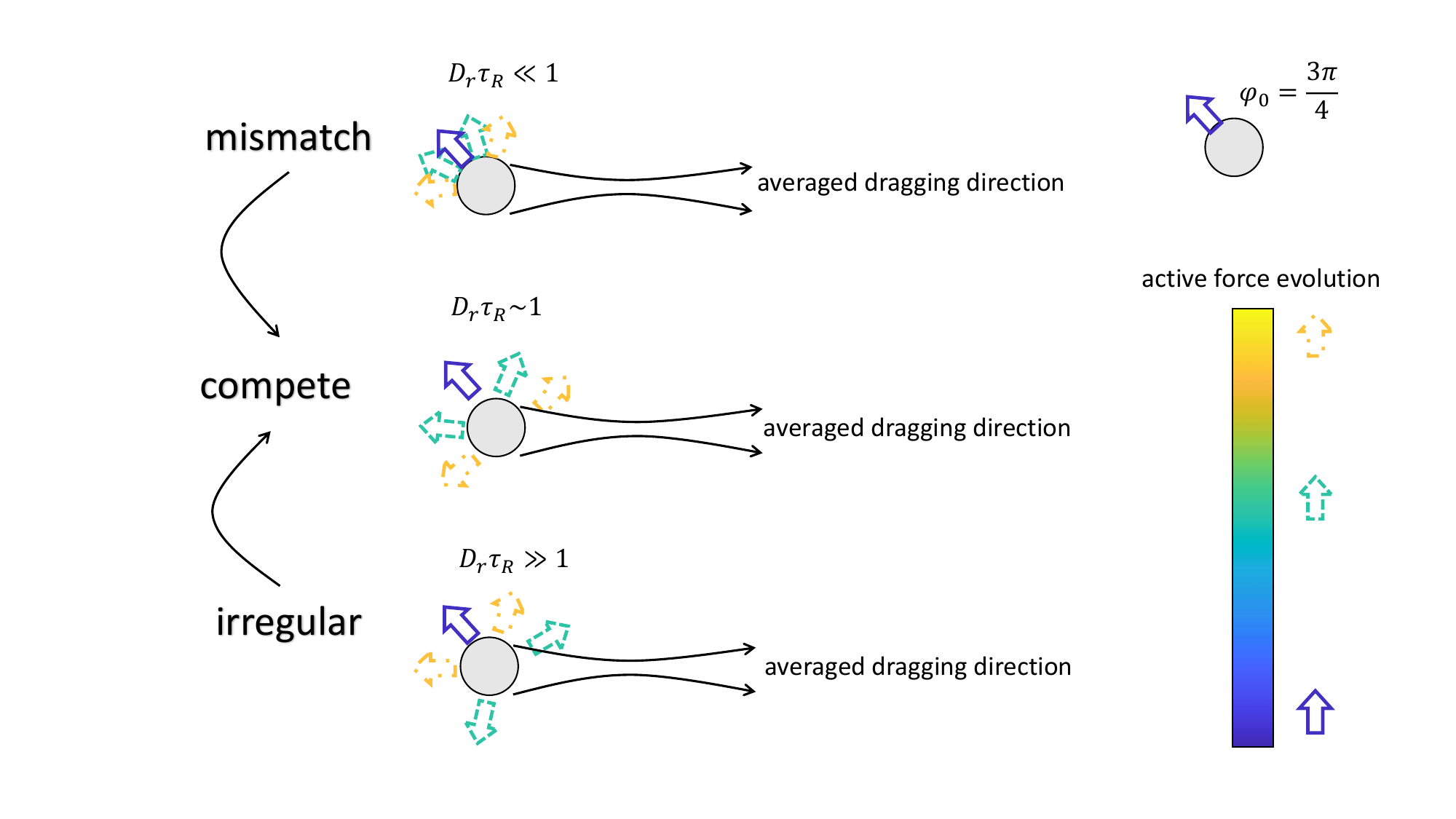}
\caption{Mechanism behind the nonmonotonic dependence on rotational diffusion $D_r$. The gray ball represents active matter, and the arrow indicates the active-force direction, with its time evolution marked by the color bar. For $D_r\tau_R\ll 1$, the active force remains close to its initially misaligned direction and creates a persistent mismatch with the desired dragging direction. For $D_r\tau_R\sim 1$, decorrelation weakens this mismatch while leaving enough predictable active forcing for extraction. For $D_r\tau_R\gg 1$, the force changes too rapidly to provide a coherent active resource.}
\label{fig:dr_mechanism}
\end{figure*}

The physical origin of the nonmonotonic $D_r$ dependence in
Fig.~\ref{fig:aoup}(a) can be seen directly from Eq.~\eqref{eq:W_orientation}.
For the parameters in Fig.~\ref{fig:aoup}(a), the persistent active force initially points up and to the left, partly against the desired transport direction. Equation~\eqref{eq:W_orientation} becomes
\begin{equation}
 W_{\rm th}(D_r)=
 \frac{1}{12}\left[1+\frac{R_1(D_r)}{\sqrt{2}}
 +\frac{R_1(D_r)^2}{4}\right]
 -\frac{R_2(D_r)}{4}.
 \label{eq:W_Dr_explicit}
\end{equation}
Both $R_1$ and $R_2$ decrease as $D_r$ increases. The decrease of $R_1$ lowers the positive mismatch term: rotational diffusion erases the long-lived active displacement that would otherwise drag the particle away from the target direction. This makes the work drop as $D_r$ is raised from the persistent regime. However, the decrease of $R_2$ weakens the negative extraction term: when the active orientation decorrelates too rapidly, little predictable active energy remains for the protocol to harvest. At large $D_r$, both $R_1$ and $R_2$ are small and the work approaches the passive finite-time dragging cost. The minimum in Fig.~\ref{fig:aoup}(a) is therefore a balance between removing a misaligned persistent bias and retaining enough active memory to extract work. If $\varphi_0$ were aligned with the target direction, the low-$D_r$ part of the curve would change because the $R_1$ term would help, rather than hinder, transport. The schematic summary in Fig.~\ref{fig:dr_mechanism} shows this crossover: the active force is too persistent at small $D_r\tau_R$, most useful at intermediate $D_r\tau_R$, and too irregular at large $D_r\tau_R$.

This interpretation also clarifies how the present calculation differs from
open-loop active-particle dragging. In the stationary open-loop AOUP problem,
the controller does not use a measured active force, so the mean active drive
is zero and the minimum-work protocol is passive-like \cite{Schuttler2025}.
Here the controller uses the conditioned force $\boldsymbol F_a(t|\mathcal I)$,
so the protocol is biased by available information, as in the force-harvesting
framework of Ref.~\cite{Olsen2025}.

\section{Conclusion and discussion}\label{sec:conclusion}

In this work, we have derived finite-time minimum-work protocols for an underdamped Brownian particle transported by a harmonic trap in the presence of a reproducible active force. This gives the optimal velocity mode $g(t)=\phi+f_a(t)/(2\gamma)$ and the smooth protocol in Eq.~\eqref{eq:lambdastar_explicit}, supplemented by the usual underdamped endpoint jumps and ideal impulses. Inertia therefore changes the protocol that realizes the optimum through the $m\dot g/k$ correction and the endpoint impulses, but it does not change the minimum mean work itself.

The same mode $g(t)$ controls the thermal work fluctuations of the minimum-work protocol. For a fixed initial microstate, the stochastic work is the thermal noise projected onto this mode, giving Eq.~\eqref{eq:var_star}. Thus the mean-work optimum and its thermal variance share the same mass independence, even though the implementing protocol remains mass dependent. This separation clarifies why the underdamped boundary operations are not merely formal: they are needed for the Green-function projection to reduce to the same mode that determines the optimal mean path.

For active-particle applications, the useful object is the conditional mean active force available to the controller. In the AOUP model this gives the explicit mean force in Eq.~\eqref{eq:mean_orientation} and the work formula Eq.~\eqref{eq:W_orientation}. The two integrals $R_1(D_r)$ and $R_2(D_r)$ separate the displacement mismatch from the active resource. For the misaligned initial orientation used in Fig.~\ref{fig:aoup}(a), increasing $D_r$ first removes a persistent bias against the dragging direction and then destroys the predictable active forcing needed for extraction, producing the observed nonmonotonic work curve. Residual active fluctuations add the covariance term in Eq.~\eqref{eq:active_var}, which accounts for the work-variance behavior in Fig.~\ref{fig:aoup}(b).

These results suggest practical strategies for measurement-informed transport and energy harvesting at microscales. In low-damping trapping platforms, such as levitated particles or active granular particles, an estimate of the initial active force or propulsion direction could be used to program a trap trajectory that most efficiently moves the particle. The mass independence of the optimal work and its conditional variance implies a common energetic performance bound across different inertial regimes, but Fig.~\ref{fig:aoup}(c,d) shows that it does not imply a common level of experimental accessibility. Mass changes both the distance traveled by the regular trap protocol and the strength of its endpoint impulses, and can therefore change the actuator bandwidth, range, and peak-force requirements by a large amount without changing either of the reported work moments.

Several extensions are natural. Real traps cannot implement instantaneous jumps or ideal delta impulses, so finite bandwidth and amplitude constraints should regularize the protocol and may reintroduce mass dependence into the achievable work itself. The geometric costs $X$ and $Y$ provide a first proxy for these limitations, while a fully constrained optimal-control calculation should impose the actual bandwidth, travel, and peak-force bounds of a chosen apparatus. Measurement error should also be included explicitly when the active force is used as conditioned information. Finally, energetic performance, implementation cost, and work stability are distinct objectives; the exact solution here provides a baseline for comparing those constrained and experimentally realistic variants.

\section*{Acknowledgments}

R. B. conceived and designed the research, proved that the work fluctuation is independent of the inertia, and co-supervised the project. K. Y. performed the research and drafted the initial manuscript. Z. H. supervised the project. All authors contributed to the discussion and revision of the manuscript. 

K.Y. and Z.H. are supported by NSFC(22533005). 
R.B. is supported by JSPS KAKENHI Grant No. 25KJ0766. 
We thank Mingrui Zhao, Yijia Cheng and Jiaan He for their helpful discussions and comments.
K. Y. thanks XinYuan Xie for computational support.

\section*{Competing Interests}

The authors declare no competing interests.

\section*{Data Availability}

The numerical data supporting the findings of this study can be reproduced using the code provided in the repository specified in the Code Availability section. No additional datasets were used in this study.

\section*{Code Availability}

The code supporting this study is available at \url{https://github.com/kangcanyao/code-of-Finite-time-minimum-work-protocols}.

\appendix

\section{Mean-work functional and endpoint structure}\label{app:mean}

Starting from Eq.~\eqref{eq:work}, set $y=\lambda-q$. Then
\begin{equation}
 \langle W\rangle=k\int_0^{t_f}(\dot y+\dot q)y\dd t
 =\left[\frac{k}{2}y^2\right]_0^{t_f}
 +k\int_0^{t_f}y\dot q\dd t .
\end{equation}
Using Eq.~\eqref{eq:mean_dyn}, $ky=m\ddot q+\gamma\dot q-f_a$, so
\begin{equation}
 \langle W\rangle=
 \left[\frac{k}{2}(\lambda-q)^2\right]_0^{t_f}
 +\int_0^{t_f}(m\ddot q+\gamma\dot q-f_a)\dot q\dd t .
\end{equation}
Integrating $m\dot q\ddot q$ gives Eq.~\eqref{eq:mean_functional}. The bulk Lagrangian is $\gamma\dot q^{\,2}-f_a\dot q$, which yields the Euler--Lagrange equation used in Sec.~\ref{sec:model_mean}. The final transversality condition gives $\lambda_f-q_*(t_f)=2\gamma\phi/k$, which fixes Eq.~\eqref{eq:phi}. Substituting $q_*$ into Eq.~\eqref{eq:mean_dyn} gives the open-interval protocol Eq.~\eqref{eq:lambdastar_explicit}.

For the distributional underdamped convention of Ref.~\cite{GomezMarin2008},
the endpoint jumps follow from the one-sided limits of
$\lambda_*^{\rm s}$ in Eq.~\eqref{eq:full_protocol}. The Dirac peaks cancel
the one-sided optimal velocities in the distributional equation of motion.

\section{Backward-kernel identity for the minimum-mean-work protocol}\label{app:kernel}

The backward kernel in Eq.~\eqref{eq:Psi_def} solves the adjoint equation
\begin{equation}
 (m\partial_t^2-\gamma\partial_t+k)\Psi(t)=k\dot\lambda_*(t)
\end{equation}
on the open interval, with final boundary conditions fixed by the singular terms in the protocol. For the smooth part of the optimum,
\begin{equation}
 \lambda_*^{\rm s}(t)=q_*(t)+\frac{\gamma g(t)}{k}
 +\frac{m\dot g(t)}{k}-\frac{f_a(t)}{k},
 \qquad \dot q_*(t)=g(t),
\end{equation}
and $g(t)=\phi+f_a(t)/(2\gamma)$. Differentiating and using $\dot f_a=2\gamma\dot g$ gives
\begin{equation}
 k\dot\lambda_*^{\rm s}(t)=m\ddot g(t)-\gamma\dot g(t)+kg(t),
 \qquad 0<t<t_f.
\end{equation}
Thus $g(t)$ solves the adjoint equation in the bulk. When $g(t)$ is cut off at $t_f$, the distributional derivatives generate boundary terms. These are precisely matched by the final jump and final delta peak in Eq.~\eqref{eq:full_protocol}; the initial singularity is handled analogously in the causal Green-function representation. By uniqueness of the adjoint boundary-value problem, $\Psi(t)=g(t)$ for $0<t<t_f$.

\section{Mass-dependent protocol cost and boundary transition}\label{app:implementation}

This appendix gives the construction used in Fig.~\ref{fig:aoup}(c,d).  The
regular, non-Dirac part of the vector protocol can be written as an affine
function of mass,
\begin{equation}
 \boldsymbol\lambda_{\rm reg}(t;m)
 =\boldsymbol\lambda_0(t)+m\boldsymbol h(t),
 \qquad
 \boldsymbol h(t)=\frac{\dot{\boldsymbol g}(t)}{k},
 \label{eq:lambda_affine_mass}
\end{equation}
where $\boldsymbol\lambda_0$ contains all mass-independent terms.  To count
both smooth motion and the finite endpoint jumps, the numerical path is
partitioned into consecutive segments.  The displacement of segment $i$ has
the affine form
\begin{equation}
 \Delta\boldsymbol\lambda_i(m)=\boldsymbol a_i+m\boldsymbol b_i,
 \qquad
 X(m)=\sum_i|\boldsymbol a_i+m\boldsymbol b_i|.
 \label{eq:X_discrete}
\end{equation}
The continuum limit gives the total variation of the regular protocol.  This
definition excludes the two Dirac peaks, whose vector coefficients follow
from Eq.~\eqref{eq:full_protocol}:
\begin{equation}
 \boldsymbol A_0=\frac{m}{k}\boldsymbol g(0),
 \qquad
 \boldsymbol A_f=-\frac{m}{k}\boldsymbol g(t_f).
 \label{eq:impulse_coefficients}
\end{equation}
Consequently,
\begin{equation}
 Y(m)=C_\delta m,
 \qquad
 C_\delta=\frac{|\boldsymbol g(0)|+|\boldsymbol g(t_f)|}
 {k\tau_R}.
 \label{eq:Y_linear}
\end{equation}
The factor $\tau_R$ is the reference time used in Fig.~\ref{fig:aoup}; a
different reference time only rescales $\kappa$.

Equation~\eqref{eq:X_discrete} is convex in $m$, since it is a sum of norms of
affine functions. Away from a vanishing segment its derivatives are
\begin{equation}
 X'(m)=\sum_i
 \frac{\boldsymbol b_i\cdot
 (\boldsymbol a_i+m\boldsymbol b_i)}
 {|\boldsymbol a_i+m\boldsymbol b_i|}.
 \label{eq:X_prime}
\end{equation}
\begin{equation}
 X''(m)=\sum_i\left[
 \frac{|\boldsymbol b_i|^2}{|\boldsymbol v_i|}
 -\frac{(\boldsymbol b_i\cdot\boldsymbol v_i)^2}
 {|\boldsymbol v_i|^3}\right]\geq0,
 \qquad
 \boldsymbol v_i=\boldsymbol a_i+m\boldsymbol b_i.
 \label{eq:X_second}
\end{equation}
We then minimize
\begin{equation}
 J_\kappa(m)=X(m)+\kappa C_\delta m,
 \qquad m\geq0.
 \label{eq:J_kappa}
\end{equation}
If $X'(0)<0$, the minimum lies at finite mass for a sufficiently weak impulse
penalty and obeys
\begin{equation}
 X'(m^*)+\kappa C_\delta=0.
 \label{eq:stationary_mass}
\end{equation}
The stationary point reaches the constraint boundary at
\begin{equation}
 \begin{aligned}
 \kappa_c&=-\frac{X'(0)}{C_\delta},\\
 m^*&>0\quad \text{for }\kappa<\kappa_c,
 \qquad
 m^*=0\quad \text{for }\kappa\geq\kappa_c.
 \end{aligned}
 \label{eq:kappa_critical}
\end{equation}
If $X'(0)\geq0$, the boundary solution is optimal for every $\kappa\geq0$.
Expanding Eq.~\eqref{eq:stationary_mass} about the critical point gives
\begin{equation}
 m^*(\kappa)=\frac{C_\delta}{X''(0)}
 (\kappa_c-\kappa)+O[(\kappa_c-\kappa)^2],
 \qquad \kappa\to\kappa_c^-.
 \label{eq:critical_scaling}
\end{equation}
Thus the optimal mass vanishes continuously with exponent one.  Moreover, the
envelope theorem gives $\dd J_\kappa^*/\dd\kappa=Y(m^*)$, which is continuous
at $\kappa_c$, whereas the second derivative changes from
$-C_\delta^2/X''(0)$ below the boundary to zero above it.  This nonanalyticity
motivates the transition-like language, but it is generated by constrained
protocol optimization and is not a thermodynamic phase transition.  Repeating
Eq.~\eqref{eq:kappa_critical} for each initial orientation produces the curve
$\kappa_c(\varphi_0)$ in Fig.~\ref{fig:aoup}(d).

\end{document}